\documentstyle[aps,amsmath,prl,preprint]{revtex}
\begin{document}
\draft
\tightenlines

\title{Self-Consistent Theory of Dynamic Melting of a Vortex Lattice}
\author{Staffan Grundberg and J{\o}rgen Rammer}
\address{Department of Theoretical Physics, Ume{\aa} University, S-901
  87 Ume{\aa}, Sweden}
\date{\today}
\maketitle
\begin{abstract}
  The dynamic melting of vortex lattices in type II superconductors is
  considered. A field-theoretic formulation of the pinning problem
  allows the average over the quenched disorder to be performed
  exactly. A self-consistent theory is constructed using a functional
  method for the effective action, allowing a determination of the
  pinning force and the vortex fluctuations.  The phase diagram for
  the dynamic melting transition is determined numerically. In
  contrast to perturbation theory, the self-consistent theory is in
  quantitative agreement with the prediction of a recent
  phenomenological theory and simulations and experimental data.
\end{abstract}

The advent of high temperature superconductors has stimulated a
renewed interest in the dynamics of vortices in type II
superconductors. In this letter we consider the influence of quenched
disorder on the dynamic melting of a vortex lattice.  This
non-equilibrium phase transition has been studied experimentally
\cite{experimentalrefs} as well as through numerical
simulation and a phenomenological theory and perturbation theory
\cite{Koshelev} - \cite{S&V2}. The notion of dynamic melting refers to
the melting of a moving vortex lattice where in addition to the
thermal fluctuations, fluctuations in vortex positions are induced by
the disorder.  A temperature-dependent critical velocity distinguishes
a transition between a phase where the vortices form a moving lattice,
the solid phase, and a vortex liquid phase.

We consider a two-dimensional system (normal to $\hat{\bf n}$) since
we have a thin superconducting film in mind, or a 3D layered
superconductor with uncorrelated disorder between the layers.
The description of the vortex dynamics will be based on the Langevin
equation \cite{Koshelev}, \cite{Blatter}
\begin{equation}
  \label{eq:langevin}
 \eta
    \dot{\bf u}_{{\bf R}t} + \sum_{{\bf R}'}
  \Phi_{{\bf R}{\bf R}'} {\bf u}_{{\bf R}'t}
  =  {\bf F} - \nabla V({\bf R}+{\bf u}_{{\bf R}t})
   +
  \boldsymbol{\xi}_{{\bf R}t},
\end{equation}
where ${\bf u}_{{\bf R}t}$ is the displacement at time $t$ of the
vortex which initially has equilibrium position ${\bf R}$, and $\eta$ is
the friction coefficient (per unit length of the vortex).  The dynamic
matrix, $\Phi_{{\bf R}{\bf R}'}$, of the triangular Abrikosov vortex
lattice describes the harmonic interaction between the vortices, and
is specified within the continuum theory of elastic media by the
compression modulus, $c_{11}$, and the shear modulus, $c_{66}$,
\cite{Brandt}
\begin{equation}
  \Phi_{\bf q} = \frac{\phi_0}{B} \left(
  \begin{array}{cc}
    c_{11} q_x^2 + c_{66} q_y^2 & (c_{11} - c_{66}) q_xq_y \\
    (c_{11} - c_{66}) q_xq_y & c_{66} q_x^2 + c_{11} q_y^2
  \end{array}
  \right),
\end{equation}
where $\phi_0 = h/2e$ is the flux quantum, and $B$ the magnitude of
the external magnetic field, $ {\bf B}=B \hat{{\bf n}}$, and
$\phi_0/B$ is therefore equal to the area, $a^2$, of the unit cell of
the vortex lattice.  The force (per unit length) on the right hand
side of eq.  (\ref{eq:langevin}) consists of the Lorentz force, ${\bf
  F} = \phi_0 \, {\bf j} \times \hat{\bf n}$, due to the (assumed
constant) transport current density $ {\bf j}$, and the thermal white
noise stochastic force, $\boldsymbol{\xi}_{{\bf R}t}$, is specified
according to the fluctuation-dissipation theorem $\langle \xi_{{\bf
    R}t}^{\alpha} \xi_{{\bf R}'t'}^{ \alpha '} \rangle = 2 \eta T
\delta(t-t') \delta_{\alpha \alpha '} \delta_{{\bf R}{\bf R}'}$, and $V$
is the pinning potential due to quenched disorder.  The pinning is
described by a Gaussian distributed stochastic potential with zero
mean, and thus characterized by its correlation function $\langle
V({\bf x}) V({\bf x}') \rangle = \nu({\bf x} - {\bf x}') = \nu_0/(2\pi
r_p^2) \exp\{-|{\bf x} - {\bf x}'|^2/(2r_p^2)\}$, taken to be a
Gaussian function with range $r_p$ and strength $\nu_0$ in our
numerical calculations.

The average vortex motion is conveniently described by reformulating
the stochastic problem in terms of the dynamic field
theory\cite{Janssen}. The probability functional for a realization
$\{{\bf u}_{{\bf R}t}\}_{{\bf R}}$ of the motion of the vortex lattice
is expressed, using the equation of motion, as a functional
integral over a set of auxiliary variables $\{\tilde{\bf u}_{{\bf
    R}t}\}_{{\bf R}} $, and we are led to consider the generating
functional
\begin{equation}
  {\cal Z}[{\bf F},{\bf J}] = \int \! \prod_{\bf R} {\cal D}{\bf
    u}_{{\bf R}t} \! \int \! \prod_{\bf R'} {\cal D} \tilde{\bf u}_{{\bf
    R}'t'}\;  e^{i{\cal S}[{\bf u},\tilde{\bf u}]},
\end{equation}
where in the action, ${\cal S}[{\bf u},\tilde{\bf u}] = \tilde{\bf u}
(D_R^{-1} {\bf u} + {\bf F} - \nabla V + \boldsymbol{\xi}) + {\bf
  J}{\bf u}$, matrix notation is used in order to
suppress the integrations over time and summations over vortex
positions and Cartesian indices.  The retarded Green's operator is
given by $ - D_R^{-1} {\bf u} = \eta \dot{\bf u}_{{\bf R}t} +
\sum_{\bf R'} {\Phi}_{\bf RR'} {\bf u}_{{\bf R}'t}, $ and its Fourier
transform is the matrix in Cartesian space $ D_R^{-1}\!({\bf q},\omega) =
i\eta \omega 1 - {\Phi}_{\bf q}$. The average with respect to
thermal noise and quenched disorder is immediately performed
and we obtain the averaged generating functional
\begin{equation}
  Z[f,K] = \langle\!\langle {\cal Z} \rangle\!\rangle
  =
  \int \! {\cal D}\phi \, e^{ iS[\phi] + if
  \phi + \frac{i}{2} \phi K \phi} ,
\end{equation}
where in order to obtain a self-consistent equation for the two-point
Green's function, we have added a two-particle source term $K$.  We
have introduced the notation $\phi = (\tilde{\bf u}, {\bf u})$ and
$f=({\bf F},{\bf J})$, and the source term, ${\bf J}$, coupling to
the vortex positions, ${\bf u}$, allow us to generate the vortex
correlation functions, say,
\begin{equation}
     \langle\!\langle {\bf u}_{{\bf R}t} {\bf u}_{{\bf R}'t'}
    \rangle\!\rangle
    = - \left.
  \frac{\delta^2 Z}{\delta {\bf J}_{{\bf R}t} \delta {\bf J}_{{\bf R}'t'}}
   \right|_{K=0,{\bf J} = {\bf 0}} .
\end{equation}
The action upon averaging, $S = S_0 + S_V$, consists of a
quadratic term, $S_0[\phi] = \phi D^{-1} \phi/2$, specified in terms
of the free inverse symmetric matrix Green's function (where the 
matrix $D^{-1}$ in the dynamical, or Keldysh, indices in addition is a
matrix in Cartesian indices, and time and vortex positions)
\begin{equation}
  D^{-1} = \left(
  \begin{array}[c]{cc}
    2i\eta T \delta_{\alpha\beta}  \delta_{{\bf R}{\bf R}'}
    \delta(t-t') & D_R^{-1} \\
    D_A^{-1} & 0
  \end{array}
  \right),
\end{equation}
and a term originating from the disorder
\begin{eqnarray}
  S_V\![\phi]\!=\!\frac{-i}{2}\!\sum_{{\bf R}{\bf R}'}
  \int_{\!-\infty}^{\!\infty}\!\!\!\!\!\!\!\!dt\, \int_{\!-\infty}^{\!\infty}
  \!\!\!\!\!\!\!\!dt' \tilde{u}_{{\bf R}t}^{\alpha}
   \nabla_{\!{\alpha}}\!\! \nabla_{\!{\beta}} \nu({\bf
  u}_{{\bf R}t}\!-\!{\bf u}_{{\bf R}'t'}) \tilde{u}_{{\bf
  R}'t'}^{\beta}.
\end{eqnarray}

In order to obtain an equation for the pinning force, we consider the
effective action $ \Gamma[\overline{\phi},G] = W[f,K] - f\overline{\phi} -
\overline{\phi} K \overline{\phi}/2 - i {\rm Tr}(GK)/2$, the generator of two-particle irreducible Green's
functions, i.e., the Legendre
transform of the generator of connected Green's functions, $ W[f,K] =
-i \ln Z[f,K]$, which satisfies the equations
\begin{eqnarray}
    \label{eq:motion}
  \frac{\delta \Gamma}{\delta \overline{\phi}} = -f - K
  \overline{\phi}, \ \ \ 
  \frac{\delta \Gamma}{\delta G} = - \frac{i}{2} K,
\end{eqnarray}
where $\overline{\phi}^{\alpha}_{{\bf R}t}$ is the average field, and
$G$ is the full connected two-point matrix Green's function of the
theory, and Tr denotes the trace over all variables.

In the physical problem of interest the sources $K$ and ${\bf J}$
vanish, and the full matrix Green's function has, due to the
normalization of the generating functional, $Z[{\bf F},{\bf
  J}\!=\!{\bf 0}, K\!\!=\!0] = 1$, the structure in Keldysh space
\begin{equation}
  G_{ij} = \left(
  \begin{array}[c]{cc}
    0   & G^A \\
    G^R & G^K
  \end{array}
  \right)
  = -i \left(
  \begin{array}[c]{cc}
    0 & \langle\!\langle \delta \tilde{u}^{\alpha} \, \delta
    {u}^{\beta} \rangle\!\rangle\\
    \langle\!\langle  \delta {u}^{\alpha} \, \delta
    \tilde{u}^{\beta}  \rangle\!\rangle
    & \langle\!\langle \delta {u}^{\alpha} \, \delta
    {u}^{\beta}  \rangle \! \rangle\\
  \end{array}
  \right),
\end{equation}
where $\delta {\bf u}_{{\bf R}t} = {\bf u}_{{\bf R}t} -
\langle\!\langle {\bf u}_{{\bf R}t} \rangle\!\rangle$ and $\delta
\tilde{\bf u}_{{\bf R}t} = \tilde{\bf u}_{{\bf R}t} - \langle\!\langle
\tilde{\bf u}_{{\bf R}t} \rangle\!\rangle$. The retarded Green's
function $G^R_{\alpha\beta}$  gives the
linear response to the force $ F_{\beta }$, and
$G^K_{\alpha\beta}$ is the correlation function (both matrices in
Cartesian indices as indicated).

As shown in \cite{CJT}, the effective action can be
written on the form $ \Gamma[\bar{\phi},G] = S[\bar{\phi}] +
i{\rm Tr} (D_S^{-1}G - \ln (D^{-1}G) - 1)/2 - i \ln \langle
e^{i S_{\rm int}[\bar{\phi},\psi]} \rangle_G^{\rm 2PI} $, where
$D_S^{-1} = \delta^2 S[\bar{\phi}] / \delta \bar{\phi} \delta
\bar{\phi}$, and $S_{\rm int}[\bar{\phi},\psi]$ is the part of
$S[\overline{\phi}+\psi]$ which is higher than second order in $\psi$
in the expansion around the average field $\overline{\phi}$. The
superscript ``2PI'' on the last term indicates that only two-particle
irreducible vacuum diagrams should be kept in the interaction part of the
effective action, and the subscript that propagator lines represent
$G$, i.e., the brackets with subscript $G$ denote the average $
\langle e^{i S_{\rm int}[\bar{\phi},\psi]} \rangle_G = (\det iG)^{-1/2}
\int\!\!{\cal D}\psi\; e^{i\psi G^{-1} \psi/2}e^{i S_{\rm
    int}[\bar{\phi},\psi]} $.  In order to get a closed expression for
the self-energy in terms of the two-point Green's function an
approximation is called for, and we expand the exponential and keep
only the first term, $ - i \ln \langle e^{i S_{\rm
    int}[\bar{\phi},\psi]} \rangle_G^{\rm 2PI} \simeq \langle
S_V[\bar{\phi}+\psi] \rangle_G^{\rm 2PI}$, i.e., the Hartree
approximation where diagrams with propagators connecting different
impurity correlators are neglected (which can also
be expressed as a Gaussian fluctuation corrected saddle-point approximation
\cite{KinzelbachHornerEckernWerner}).

For the problem of interest, the two-particle source, $K$, vanishes,
and the second equation in (\ref{eq:motion}) yields the Dyson
equation, $G^{-1} = D^{-1} - \Sigma[\bar{\phi},G]$, with the Keldysh matrix
self-energy given by
\begin{equation}
  \label{eq:selfenergy}
  \Sigma_{ij}
  = \left(
  \begin{array}[c]{cc}
    \Sigma^K   & \Sigma^R \\
    \Sigma^A & 0
  \end{array}
  \right)
  = 2i \left. \frac{\delta \langle S_V
  [\bar{\phi}+\psi] \rangle_G^{\rm 2PI}}{\delta G_{ij}}
\right|_{\scriptsize
  \begin{array}[b]{c}
    K=0 \\
   \langle\!\langle
\tilde{\bf u} \rangle\!\rangle = {\bf 0}
  \end{array}}. 
\end{equation}
The Dyson equation and eq. (\ref{eq:selfenergy}) constitute a set of
self-consistent equations for the Green's functions and the
self-energies. Since the expectation value of the auxiliary field
vanishes due to the normalization of the generating functional, the
average field entering eq. (\ref{eq:selfenergy}) is $\bar{\phi} =
(\langle\!\langle \tilde{\bf u}_{{\bf R}t} \rangle\!\rangle,
\langle\!\langle {\bf u}_{{\bf R}t} \rangle\!\rangle) = ({\bf 0},{\bf
  v}t)$, where ${\bf v}$ is the average lattice velocity. The matrix
self-energy has two independent components, say $\Sigma^K$ and
$\Sigma^R$, since $\Sigma_{ \beta \alpha }^A ({\bf R}t,{\bf R}'t') =
\Sigma_{ \alpha \beta}^R({\bf R}'t',{\bf R}t)$.  From eq.
(\ref{eq:selfenergy}) we obtain for the case of $N$ vortices
$\Sigma^K_{\alpha\beta}({\bf R}t,{\bf R}'t') = - i/(Na^2) \sum_{\bf k}
\nu({\bf k}) k_{\alpha} k_{\beta} e^{i\varphi_{\bf k}}$, and $
\Sigma^R({\bf q},\omega) = \sigma^R({\bf q},\omega) - \sigma^R({\bf
  q}\!=\!{\bf 0}, \omega\!=\!0)$, where the Fourier transform has the
Cartesian components $ \sigma^R_{\alpha\beta} ({\bf R}t,{\bf R}'t') =
1/(Na^2) \sum_{\bf k} \nu({\bf k}) k_{\alpha} k_{\beta} ({\bf k}
G^R({\bf R}t,{\bf R}'t') {\bf k}) e^{i\varphi_{\bf k}}$. The influence
of thermal and disorder induced fluctuations is described by the
complex phase $ \varphi_{\bf k} = i{\bf k}M{\bf k} + {\bf k} \cdot
({\bf R} - {\bf R}' + {\bf v}(t-t'))$, specified by the Cartesian
matrix $ M_{ \alpha \beta }({\bf R}t,{\bf R}'t') = i(G_{ \alpha \beta
  }^K({\bf R}t,{\bf R}t) - G_{ \alpha \beta }^K({\bf R}t,{\bf
  R}'t'))$. Using the Langevin equation and the first equation in
(\ref{eq:motion}) we obtain for the pinning force, ${\bf F}_{p} = -
\langle \! \langle \nabla V \rangle \!  \rangle$,
\begin{equation}
  \label{eq:pinningforce}
  {\bf F}_{p} = 
  i \sum_{{\bf R}'} \int\limits_{-\infty}^{\infty} \!\!\!\!dt 
 \!\! \int\limits_{}^{}\!\!\! \frac{d{\bf k}}{(2\pi)^2}\, {\bf k} \,
  \nu({\bf k}) ({\bf k} {G^R}({\bf R}t,\!{\bf R}'t'){\bf k})
  e^{i\varphi_{\bf k}}.
\end{equation}

Let us recall the argument for determining the phase diagram
for dynamic melting of a vortex lattice presented in \cite{Koshelev}.
There the disorder induced fluctuations were estimated considering the
correlation function, $\kappa_{\beta \alpha }({\bf x},t) =
\langle\!\langle f_{\beta}({\bf x},t) f_{\alpha}({\bf 0},0)
\rangle\!\rangle$, of the pinning force density of the vortices, ${\bf
  f}({\bf x},t) = - \sum_{\nu} \delta({\bf x} - {\bf
  R}_{\nu}(t))\nabla V({\bf x} - {\bf v}t)$.  Neglecting the
interdependence of the fluctuations of the vortex positions and the
fluctuations in the disorder potential, the pinning force correlation
function factorizes, and since in the fluidlike phase the motion of
different vortices are ``incoherent'', one gets $ \kappa_{ \beta
  \alpha}({\bf x},t) \simeq - n_{_V} \delta({\bf x}) \nabla_{\!\beta }
\nabla_{\!\alpha } \nu ({\bf v}t)$, $n_{_V}$ being the density of
vortices.  In analogy with the noise correlator the effect of disorder
induced fluctuations is represented by a ``shaking temperature''
\begin{equation}
  \label{eq:shaking}
 T_{{\rm sh}} = \frac{1}{4\eta n_{_V}} \sum_{\alpha}\!\!\!
  \int \!\!\! d{\bf x}\!\!\! \int \!\!\!
  dt \;\kappa_{\alpha\alpha}({\bf x},t)
  = \frac{ \nu_0}{4\sqrt{2\pi}F r_p^3},
\end{equation}
where in the last equality it is assumed that the pinning force is
small compared to the friction force, i.e., $\eta v \simeq F$. An
effective temperature is then obtained by adding the ``shaking
temperature'' due to disorder to the temperature, and according to eq.
(\ref{eq:shaking}) the effective temperature decreases with increasing
external force (i.e., with increasing average velocity of the
vortices). As the external force is increased the fluid thus freezes
into a lattice. The value of the external force for which the moving
lattice melts, the transition force $F_t$, is in \cite{Koshelev} taken
to be the value for which the effective temperature equals the melting
temperature in the absence of disorder, $ T_{\rm eff}(F\!\!=\!\!F_t) =
T_m$, and has, according to eq.
(\ref{eq:shaking}), the temperature dependence $F_t =
\nu_0/(4\sqrt{2\pi}r_p^3 (T_m-T))$, as the temperature approaches the
melting temperature of the ideal lattice from below (we note that the
transition force for strong enough disorder exceeds the critical force
for which the lattice is pinned $F_t > F_c = 0.2 \nu_0^{1/2}/r_p^2$).

We now describe the calculation of the physical quantities of interest
using the self-consistent theory. The conventional way of determining
the melting transition is to use the Lindemann criterion, which states
that a lattice melts when the displacement fluctuations reach a
critical value $\langle\langle {\bf u}^2 \rangle\rangle = c_L^2 a^2$, where $c_L$ is
the Lindemann parameter which is typically ranging in the interval
from 0.1 to 0.2. In two dimensions the position fluctuations of a
vortex diverge even for a clean system, and the Lindemann criterion
implies that a two-dimensional vortex lattice is always unstable
against thermal fluctuations. However a quasi long-range translational
order persists up to a certain melting temperature \cite{S&V}. As a
criterion for the loss of long-range translational order a modified
Lindemann criterion involving the relative vortex fluctuations,
$\langle\langle ({\bf u}({\bf R}+{\bf a}_0,t) - {\bf u}({\bf R},t))^2 \rangle\rangle
= 2c_L^2 a^2$, where ${\bf a}_0$ is a primitive lattice vector, has
successfully been employed \cite{S&V}, and its validity verified
within a variational treatment \cite{Nattermann}. The relative
displacement fluctuations are specified in terms of the correlation
function, $\langle\!\langle ({\bf u}({\bf R}+{\bf a}_0,t) - {\bf
  u}({\bf R},t))^2 \rangle\!\rangle = 2i {\rm tr} \! \left( G^K({\bf
  0},0) - G^K({\bf a}_0,0) \right)$, where tr denotes the trace with
respect to the Cartesian indices. The correlation function is
according to the Dyson equation $ G^K_{{\bf q}\omega} = G^R_{{\bf
    q}\omega} (\Sigma^K_{{\bf q}\omega} - 2i\eta T) G^A_{{\bf
    q}\omega}$, where the influence of the quenched disorder appears
explicitly through $\Sigma^K$ and implicitly through $\Sigma^R$ and
$\Sigma^A$ in the advanced and retarded response functions.
Furthermore, the self-energies depend self-consistently on the
response and correlation functions. We have calculated numerically the
Green's functions and self-energies and thereby the vortex
fluctuations for a vortex lattice of size $8 \times 8$, and evaluated
the pinning force from eq.  (\ref{eq:pinningforce}).

We determine the phase diagram for dynamic melting of the vortex
lattice by calculating the relative displacement fluctuations for a
set of velocities, and interpolate to find the transition velocity,
$v_t$, i.e., the value of the velocity at which the fluctuations
fulfill the modified Lindemann criterion. The transition force is
given by the averaged equation of motion, $F_t = \eta v_t + F_p(v_t)$,
and is then determined by using the numerically calculated pinning
force.  Repeating this procedure for various temperatures determines
the melting curve, i.e., the temperature dependence of the transition
force, $F_t(T)$, separating two phases in the $FT$-plane: a high
velocity phase where the vortices form a moving solid when the
external force exceeds the transition force $F>F_t(T)$, and a liquid
phase for forces below the transition force.

In order to be able to compare the results of the self-consistent
theory to the simulation results in \cite{Koshelev}, we choose the
same parameters. There the melting temperature in the absence of
disorder is taken to be $ T_m=0.007$ ($2(\phi_0/4\pi \lambda )^2$
is chosen as the unit of energy per unit length, where $ \lambda $ is
the London penetration depth), the value obtained by simulations of
clean systems, and assumed equal to the Kosterlitz-Thouless
temperature, $ T_{\rm KT} = c_{66}a^2/ 4\pi$, determining thereby the shear
modulus to have the value $c_{66} = 0.088$ ($a$ is chosen as the
unit of length) \cite{Kosterlitz}.  The London penetration depth is
according to \cite{S&V} taken to equal the range of the vortex
interaction potential used in \cite{Koshelev}, which is approximately
equal to the lattice spacing, $a_0$, giving for the compression
modulus $c_{11} = 16 \pi \lambda^2 c_{66}/ a_0^2 \simeq 50 c_{66}
\simeq 4.4$ \cite{Blatter}.  The range and strength of the disorder
correlator in the simulations are $r_p = 0.2$ and $\nu_0 = 1.42 \cdot
10^{-5}$ in the chosen units.

Once the Lindemann parameter is determined, our numerical results for
the relative displacement fluctuations can be used to obtain the
dynamic phase diagram. In order to do so we calculate ``melting
curves,'' as described above, using the self-consistent theory for a
set of different values of the Lindemann parameter. We find that these
curves have the same shape, close to the melting temperature, as the
melting curve obtained from the shaking theory, $ T = C_1 - C_2/F_t$.
The curve which intersects closest to the melting temperature
$T_m=0.007$, the one depicted in the inset in the figure, then
determines the Lindemann parameter to be given by the value $c_L^2 =
0.0153$. The corresponding phase diagram obtained from the
self-consistent theory is shown in the figure, as well as the
simulation data in \cite{Koshelev}, and the melting curve obtained
from the shaking theory. The melting of the vortex lattice was in the
simulations indicated by an abrupt increase in the structural
disorder, a different melting criterion, and the agreement of the
self-consistent theory with the simulation data are therefore
independently validating the use of the modified Lindemann criterion.
The shaking theory is seen to be in remarkably good agreement with the
self-consistent theory even at lower velocities where the shaking
argument is not \emph{a priori} valid, a feature which, however, is
less pronounced for stronger disorder. It is also of interest to
recall, that while the melting curve obtained from the shaking theory
was based on arguments only valid in the liquid phase, i.e., freezing
of the vortex liquid was considered, the melting curve we obtained
from the self-consistent theory is calculated assuming a lattice,
i.e., we consider melting of the moving lattice. As apparent from the
inset in the figure, the critical exponent obtained from the
self-consistent theory, $1.0$, is in excellent agreement with the
prediction of the shaking theory, where the critical exponent equals
one.  Furthermore, we find that the self-consistent theory yields the
value $1.65 \cdot 10^{-4}$ for the constant $C_2$, a value fairly
close to the one predicted by the shaking theory, $\nu_0 /
(4\sqrt{2\pi}r_p^3) = 1.77 \cdot 10^{-4}$.

The melting curve predicted by lowest order perturbation theory is
also shown in the figure. The correlation function is in this case
given by $G^{K(1)}_{{\bf q}\omega} = D^R_{{\bf q}\omega}
[\Sigma^{K(1)}_{{\bf q}\omega} - 2i\eta T - 2i\eta T (
\Sigma^{R(1)}_{{\bf q}\omega} D^R_{{\bf q}\omega} + D^A_{{\bf
    q}\omega} \Sigma^{A(1)}_{{\bf q}\omega})] D^A_{{\bf q}\omega}$,
where the self-energies are calculated to first order in $\nu_0$.  As to
be expected, the perturbation theory results are at high velocities in
good agreement with the self-consistent theory. However, we observe
that the melting curve obtained from lowest order perturbation theory
deviates markedly at intermediate velocities from the prediction of
the self-consistent theory, and thereby the shaking theory, which is
known to account well for the measured melting curve, Hellerqvist {\em
  et. al.} \cite{experimentalrefs}.

In conclusion, we have developed a self-consistent theory of the
dynamic melting transition of a vortex lattice, enabling us to
determine numerically the melting curve directly from the dynamics of
the system. The self-consistent theory corroborates the
phase diagram obtained by the phenomenological shaking theory, 
whereas lowest order perturbation theory does not. The melting curve
obtained from the self-consistent theory is found to be in good
quantitative agreement with the phenomenological theory as well as
with simulations and experimental data.

This work was supported by the Swedish Natural Research
Council through contracts F-AA/FU 10199-314 (SG) and
F-AA/FU 10199-313 (JR).

\begin{figure}[htbp]
  \ \vspace{1cm}
  \begin{center}
    \leavevmode
    \input{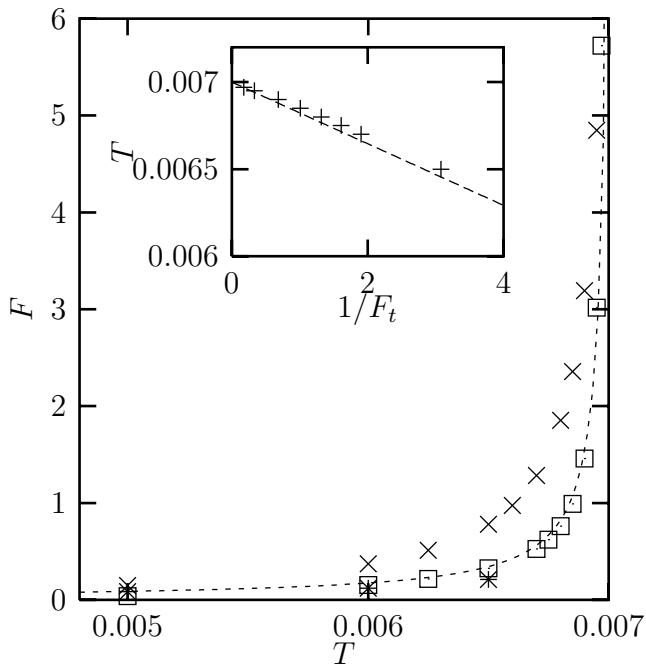}
     \caption{The dynamic melting phase diagram. The
      points on the melting curve separates the two phases - for
      forces stronger than the transition force the moving vortices
      form a solid, and for weaker forces a liquid. The dots in the
      boxes represent points on the melting curve obtained from the
      self-consistent theory for an $8 \times  8$ lattice, while
      the three stars represent the simulation results in [7].  The
      crosses represent the  results of lowest order perturbation theory.
      The dashed line is the curve $F_t(T) = 1.77 \cdot
      10^{-4}/(0.007-T)$, the melting curve obtained from the shaking
      theory. Inset: Temperature as a
      function of the inverse transition force obtained from the
      self-consistent theory (plus signs), close to the melting temperature, for
      the particular value of the Lindemann parameter $c_L^2 =
      0.0153$, for which the curve intersects the $T$-axis at $T_m
      = 0.00701$. The points lie on a straight line just as the
      prediction of the shaking theory, i.e., the dashed line $F_t
      \sim 1/(T_m - T)$, yielding the value one for the critical
      exponent.}
\label{fig:melting}
  \end{center}
\end{figure}
\end{document}